\documentclass[iop,apj,tighten,revtex4]{emulateapj}

\usepackage{multirow}
\usepackage{hhline}
\usepackage{url}
\usepackage{amssymb}
\usepackage{natbib}

\def\ltsima{$\; \buildrel < \over \sim \;$}
\def\gtsima{$\; \buildrel > \over \sim \;$}
\def\lsim{\lower.5ex\hbox{\ltsima}}
\def\gsim{\lower.5ex\hbox{\gtsima}}
\def\lapp{\ifmmode\stackrel{<}{_{\sim}}\else$\stackrel{<}{_{\sim}}$\fi}
\def\gapp{\ifmmode\stackrel{>}{_{\sim}}\else$\stackrel{<}{_{\sim}}$\fi}
\def\rmxaa{Rev. Mexicana Astron. Astrofis.}

\newdimen\minuswidth    

\setbox0=\hbox{$-$}

\minuswidth=\wd0

\catcode`@=\active

\def@{\kern\minuswidth}

\newdimen\digitwidth    

\setbox0=\hbox{\rm0}

\shorttitle{}
\shortauthors{}

\begin{document}
\title{Proper motions and structural parameters of the Galactic
  globular cluster M71\footnote {Based on observations collected with
    the NASA/ESA HST (GO10775, GO12932), obtained at the Space
    Telescope Science Institute, which is operated by AURA, Inc.,
    under NASA contract NAS5-26555.}}

\author{M. Cadelano\altaffilmark{1,2},
  E. Dalessandro\altaffilmark{1,2}, F. R. Ferraro\altaffilmark{1},
  P. Miocchi\altaffilmark{1}, B. Lanzoni\altaffilmark{1},
  C. Pallanca\altaffilmark{1} and D. Massari\altaffilmark{2,3}}
\affil{\altaffilmark{1} Dipartimento di Fisica e Astronomia,
  Universit\`a di Bologna, Viale Berti Pichat 6/2, I-40127 Bologna,
  Italy }
\affil{\altaffilmark{2} INAF - Osservatorio Astronomico di Bologna,
  Via Ranzani 1, I-40127 Bologna, Italy }
\affil{\altaffilmark{3} Kapteyn Astronomical Institute, University of
  Groningen, P.O. Box 800, 9700 AV Gr\"oningen, The Netherlands }

\begin{abstract}
By exploiting two ACS/HST datasets separated by a temporal baseline of
$\sim7$ years, we have determined the relative stellar proper motions
(providing membership) and the absolute proper motion of the Galactic
globular cluster M71.  The absolute proper motion has been used to
reconstruct the cluster orbit within a Galactic, three-component,
axisymmetric potential. M71 turns out to be in a low latitude
disk-like orbit inside the Galactic disk, further supporting the
scenario in which it lost a significant fraction of its initial mass.
Since large differential reddening is known to affect this system, we
took advantage of near-infrared, ground-based observations to
re-determine the cluster center and density profile from direct star
counts.  The new structural parameters turn out to be significantly
different from the ones quoted in the literature. In particular, M71
has a core and a half-mass radii almost 50\% larger than previously
thought.  Finally we estimate that the initial mass of M71 was likely
one order of magnitude larger than its current value, thus helping to
solve the discrepancy with the observed number of X-ray sources.
\end{abstract}


\date{18 november 2016}

\section{INTRODUCTION}
\label{intro}
Galactic globular clusters (GCs) are dense and old ($t>10$ Gyr)
stellar systems {containing up to $\sim10^6$ stars}, orbiting the Milky Way
halo and bulge. Their study is crucial to understand
the dynamical evolution of collisional systems
\citep[e.g.][]{meylan97, ferraro12} and the interplay between dynamics
and stellar evolution \citep[e.g.][]{goodman89,phinney93,
  rasio07,ferraro09b,ferraro15}. Their high central densities provide
the ideal ground to the formation of exotic objects like blue
straggler stars, cataclysmic variables, low-mass X-ray binaries and
millisecond pulsars \citep[e.g.][]{ferraro97,ferraro03,pooley03, ransom05, heinke05}.

In this respect, remarkable is the case of M71, which is a low-density
GC located at a distance of about 4 kpc from Earth. It has a quite
high metallicity ([Fe/H]$=-0.73$), a color excess $E(B-V)=0.25$
\citep[][2010 edition]{harris96} and a total mass of about $2 \times
10^4 M_\odot$ \citep{kimmig15}. X-ray observations revealed that it
hosts a large population of X-ray sources, most likely consisting of
stellar exotica.  Surprisingly, as discussed in \citet{elsner08,
  huang10}, the number of X-ray detections in M71 is significantly
larger than what is expected from its present-day mass and its
collisional parameter (which is a characteristic indicator of the
frequency of dynamical interactions and thus of the number of stellar
exotica in a GC; e.g. \citealp{bahramian13}).  However, it is worth
noticing that M71 is located at a low Galactic latitude
($l=56.75^{\circ},b=-4.56^{\circ}$), likely on a disk-like orbit
\citep{geffert00}. Hence, it could have have lost a substantial
fraction of its initial mass, due to heavy interactions with the
Galactic field and to shocks caused by encounters with molecular
clouds and/or spiral arms.  {Moreover the structural parameters 
of this cluster have been estimated from shallow optical images \citep{peterson97}, 
and therefore need to be re-determined more accurately.}
Hence, the value of the collisional parameter,
which directly depends on the cluster structural parameters
\citep{verbunt87}, could be biased.

By taking advantage of two epoch of observations {obtained with the {\it Hubble Space Telescope} (HST) }and wide-field
near-infrared and optical datasets for M71, here we present the
determination of: \emph{(i)} the stellar proper motions (which allow
us to distinguish cluster members from Galactic contaminants),
\emph{(ii)} the absolute PM of the system (from which we estimate its
orbit within the Galaxy during the last 3 Gyr), and $(iii)$ the
cluster gravitational center and structural parameters. 

In Section \ref{obs} we describe the procedures adopted for the data
reduction and analysis. Sections \ref{pm} and \ref{abspm} are devoted
to the determination of relative stellar proper motions (PMs), and of
the cluster absolute PM and orbit, respectively. In Section
\ref{clust} we present the new determination of cluster gravity
center, density profile and structural parameters from near-infrared
data and we study how the latter change if optical observations are used
instead. We also provide an estimate of the initial mass of the
system. Finally, in Section \ref{conclu} we summarize the results and discusse the
X-ray source abundance discrepancy in light of the new values of the
cluster structural parameters and the initial mass estimate.

\section{OBSERVATIONS AND DATA REDUCTION}
\label{obs}  
The present work is based on two different datasets. Their
characteristics and the adopted data reduction procedures are
described in the following.

\begin{figure}[hb]
\centering
\includegraphics[width=8.5cm]{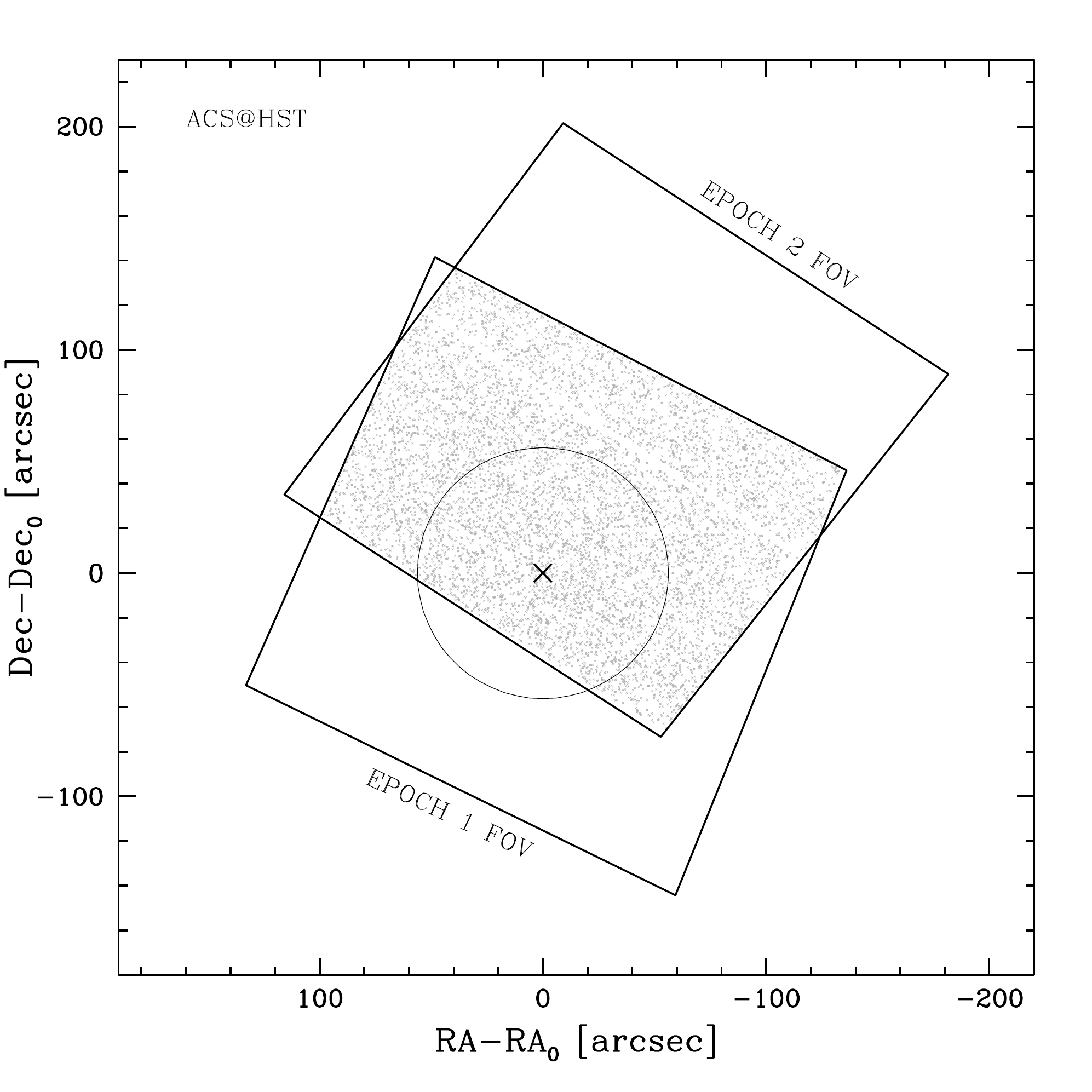}
\caption{FOVs of the ACS first and second epoch datasets, centered on
  the newly estimated gravity center of M71 (black cross; see Section
  \ref{center}). The grey dots highlight the stars in common are between
 the two datasets, which has been used to measure the stellar proper
  motions. The solid circle marks the core radius of the cluster as
  derived in this work ($r_c=56.2\arcsec$; see Section
  \ref{density}). }
\label{mappafov}
\end{figure}

\begin{figure}[bh]
\centering
\includegraphics[width=8.5cm]{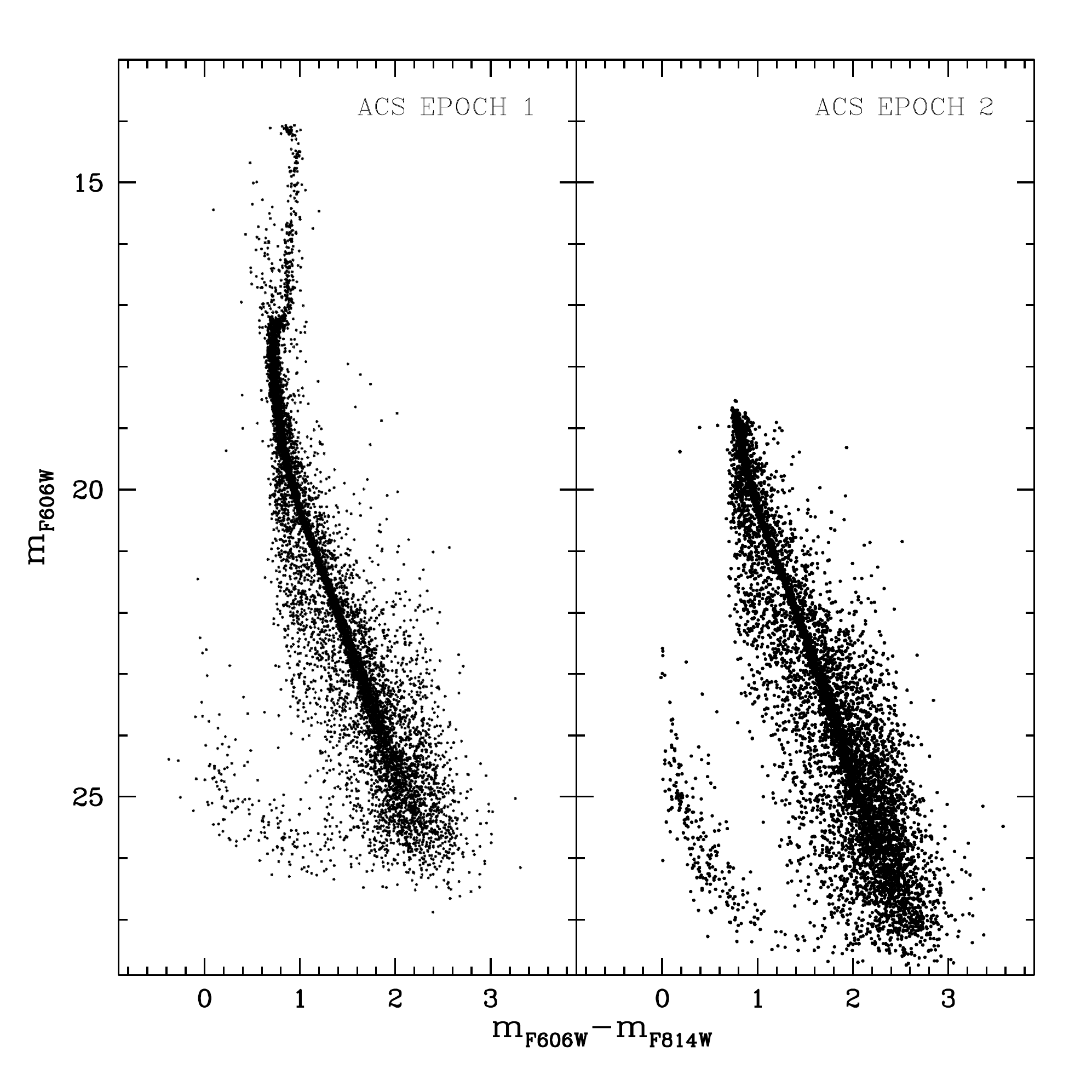}
\caption{Optical CMD of M71 obtained from the first and second epoch
  ACS datasets (left and right panels, respectively).}
  \label{cmd}
\end{figure}

{\it High-Resolution Dataset --} This has been used to determine the
stellar PMs. It consists of two sets of images acquired {with the Wide Field Channel (WFC) of the} 
Advanced Camera for Surveys (ACS) mounted on HST (see Figure \ref{mappafov} for a map of the fields of view -
FOVs - covered by these observations). This camera provides a FOV of  $202\arcsec\times202\arcsec$ with a pixel scale {of $\rm 0.05\arcsec \ pixel^{-1}$}. The first epoch data have been
collected under GO10775 (P.I.: Sarajedini) on 2006 July 1, and consist
of a set of ten dithered images, five in the F606W filter (with
exposure times: $1 \times 4$ s; $4 \times 75$ s) and five in the F814W
filter ($1 \times 4$ s; $4 \times 80$ s).  The second epoch is
composed of proprietary data obtained under GO12932 (P.I.: Ferraro) on
2013, August 20. It consists of a set of ten deep images acquired
through the F606W filter ($2 \times 459$ s; $3 \times 466$ s; $5
\times 500$ s) and nine images in the F814W filter ($5 \times 337$ s;
$3 \times 357$ s; $1 \times 440$ s). The photometric analysis has been
performed on the {\textrm -flc} images (which are corrected for flat
field, bias, dark counts and charge transfer efficiency) following the
procedures described in detail in \citet{anderson06}. Briefly, both
the epochs have been analyzed with the publicly available program
{\textrm img2xym\_WFC.09x10}, which uses a pre-determined model of a
spatially varying point spread function (PSF) plus a single
time-dependent perturbation PSF (to account for focus changes or
spacecraft breathing). {The final output of this process are two catalogs (one for each epoch)
with instrumental magnitudes and {positions for all the sources above a given threshold}.
Star positions were corrected in each catalog for geometric distortion by adopting the solution provided by \citet{anderson06}.}
By using the stars in common with the public catalog of
\citet[][see also \citealp{anderson08}]{sarajedini07}, instrumental
magnitudes have been calibrated on the VEGAMAG system and instrumental
positions have been reported on the absolute right ascension and declination coordinate
reference system ($\alpha$ and $\delta$, respectively). The final
color-magnitude diagrams (CMDs) are shown in Figure \ref{cmd} for the
two different epochs.

\begin{figure}[t]
\centering
\includegraphics[width=8.5cm]{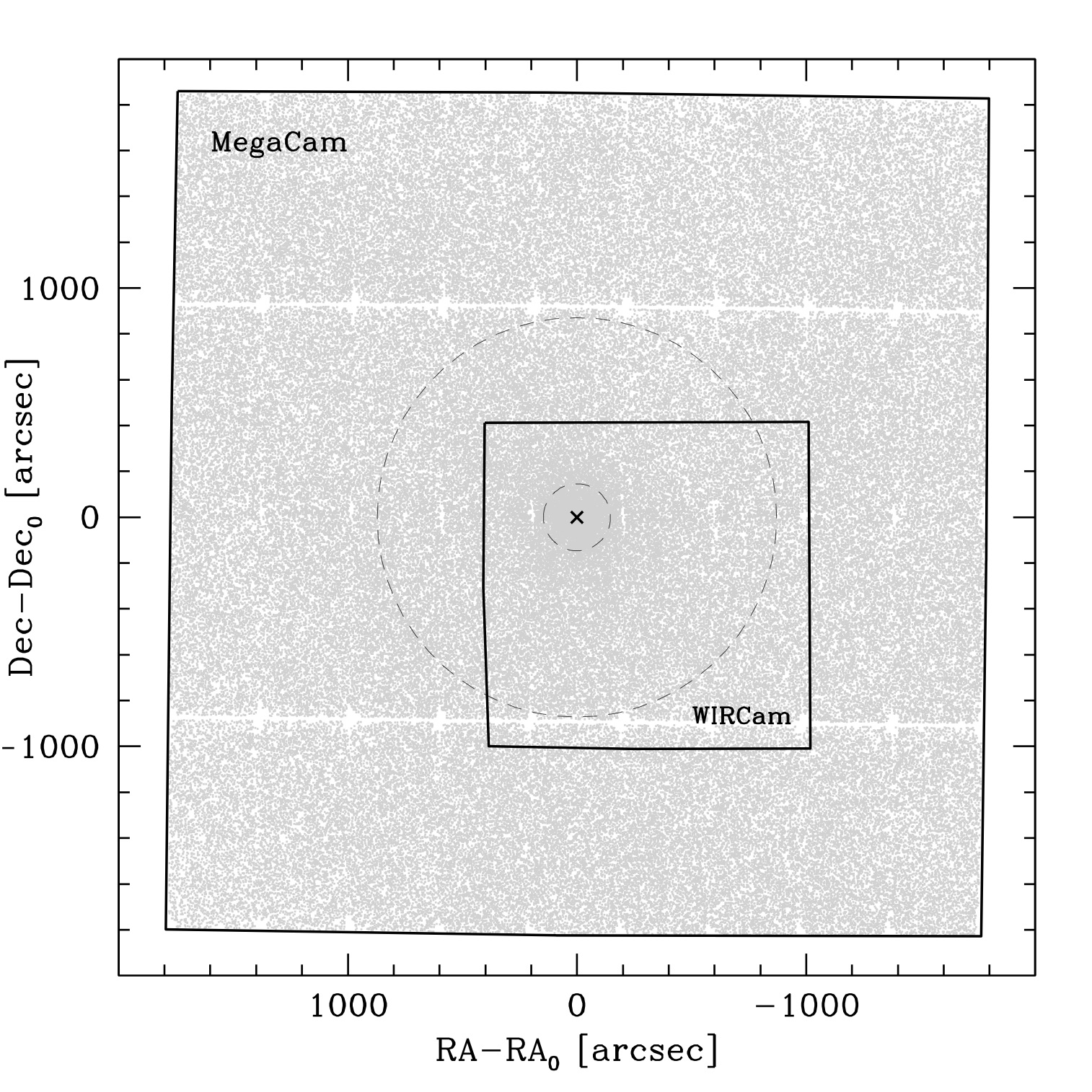}
\caption{FOVs of the MegaCam and WIRCam datasets, centered on the
  cluster gravity center (black cross). The small and large dashed
  circles mark, respectivey, the half-mass and the tidal radii derived
  in this work (see Section \ref{density}).}
\label{mapground}
\end{figure}

{\it Wide-field Dataset --} To determine the cluster gravitational
center and structural parameters, we used ground-based near-infrared
images (Prop ID: 11AD90; PI: Thanjavur) obtained with the wide field
imagers WIRCam mounted at the Canada-France-Hawaii Telescope
(CFHT). To study the effect of differential reddening, we also made
use of optical wide-field images (Prop ID: 04AC03, 03AC16; PI: Clem)
acquired with MegaCam at the same telescope.  The WIRCam camera
consists of a mosaic of four chips of 2040$\times$2040 pixels each,
with a pixel scale of $\rm 0.31\arcsec \ pixel^{-1}$, providing a
total FOV of $\sim21.5\arcmin \times 21.5\arcmin$. We analyzed seven
images obtained with the $J$ and $K_s$ filters, with exposure times of 5 s
and 24 s, respectively. A dither pattern of few arcseconds was applied
to fill the gaps among the detector chips.  The MegaCam camera
consists of a mosaic of 36 chips of 2048$\times$4612 pixels each, with
a pixel scale of $\rm 0.185\arcsec \ pixel^{-1}$ providing a FOV of
$\sim1^{\circ}\times1^{\circ}$. A total of 50 images have been
acquired, both in the g' and in the r' bands, with exposure times of
250 s each. A dither pattern of few arcseconds was adopted for each
pointing, thus allowing the filling of most of the interchip gaps,
with the exception of the most prominent, horizontal ones.  Figure
\ref{mapground} shows the map of the Wide-field dataset.

\begin{figure}[!h]
\centering
 \includegraphics[width=8.5cm]{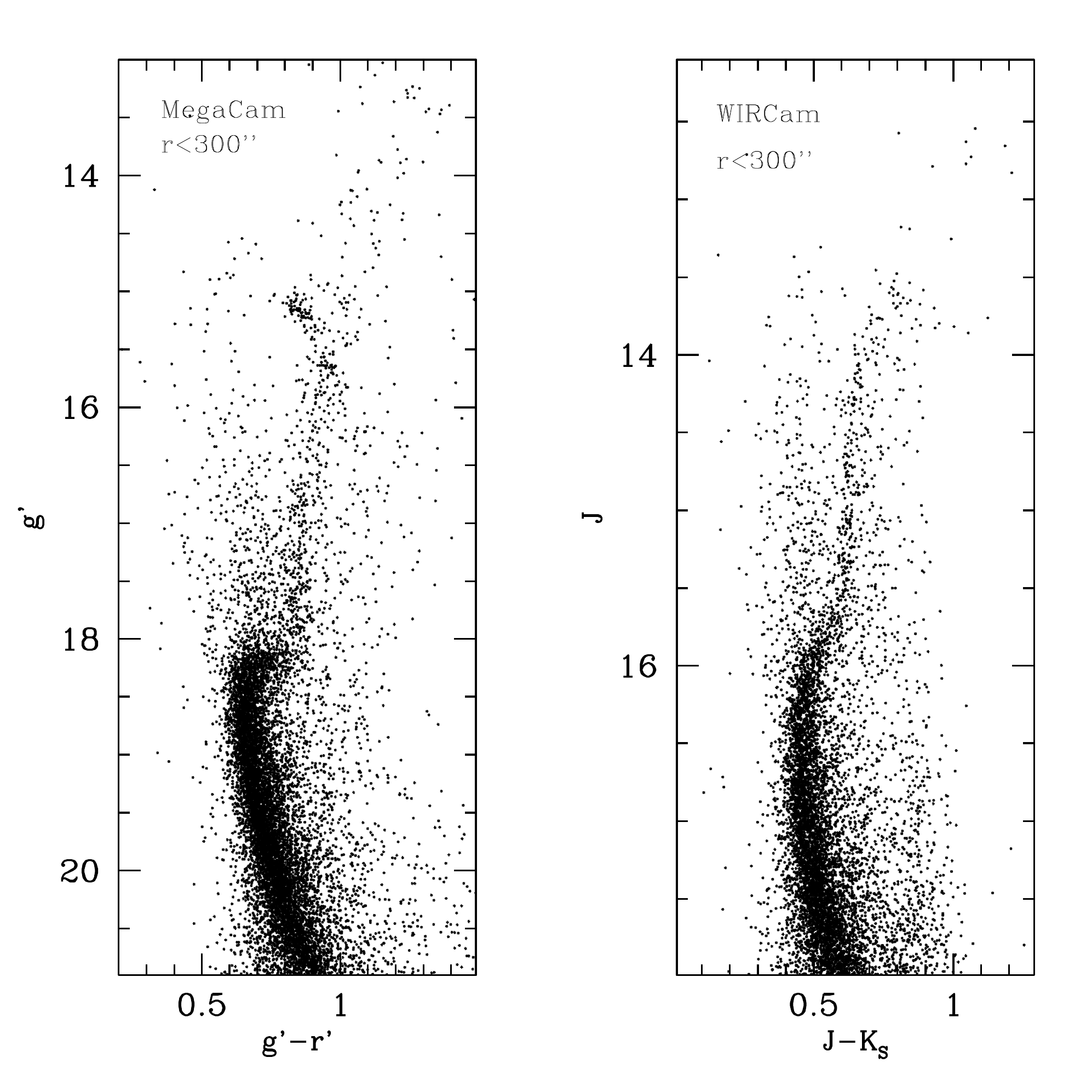}
\caption{Optical and near-infrared CMDs of M71, obtained from the
  MegaCam (left panel) and the WIRCam datasets (right panel),
  respectively.  In both cases, only stars within $300\arcsec$ from
  the center are plotted. }
\label{cmd2}
\end{figure}

For both these sets of observations, the images were pre-processed
(i.e. bias and flat-field corrected) by means of the Elixir pipeline
developed by the CFHT team and the photometric analysis has been
performed independently on each chip by following the procedures
described in \citet{dalessandro15}. Briefly, by means of an iterative
procedure, an adequate number ($> 20$) of isolated and bright stars
has been selected in each chip and filter to model the PSF. Hence, the
PSF model has been applied to all the stellar-like sources at about
4$\sigma$ from the local background by using {\textrm DAOPHOT} and the
PSF-fitting algorithm ALLSTAR \citep{stetson87}. For each filter and
chip, we matched the single-frame catalogs to obtain a master
list. Each master list includes the instrumental magnitudes, defined
as the weighted mean of the single image measurements reported to the
reference frame of the transformation, and the error, which is the
standard deviation of the mean. Instrumental magnitudes have been
reported to the SDSS photometric system\footnote{See
  \url{http://www.cfht.hawaii.edu/Science/CFHTLS-DATA/megaprimecalibration.html\#P2.}}
for the MegaCam catalog, and to the 2MASS system for the WIRCam
catalog. Finally the instrumental positions have been reported to
the absolute coordinate reference frame by using the stars in common
with the 2MASS catalog\footnote{Publicly available at
  \url{http://vizier.u-strasbg.fr}}. The CMDs for these datasets are
shown in Figure \ref{cmd2} for stars located at less than $300\arcsec$
from the center.

As can be seen from both Fig. \ref{cmd} and Fig. \ref{cmd2}, the
standard evolutionary sequences are well defined. However, they are
also heavily contaminated by foreground objects, as expected from the
location of M71 close to the Galactic disk.

\section{RELATIVE PROPER MOTIONS}
\label{pm}
\begin{figure}[t]
\centering
\includegraphics[width=8.5cm]{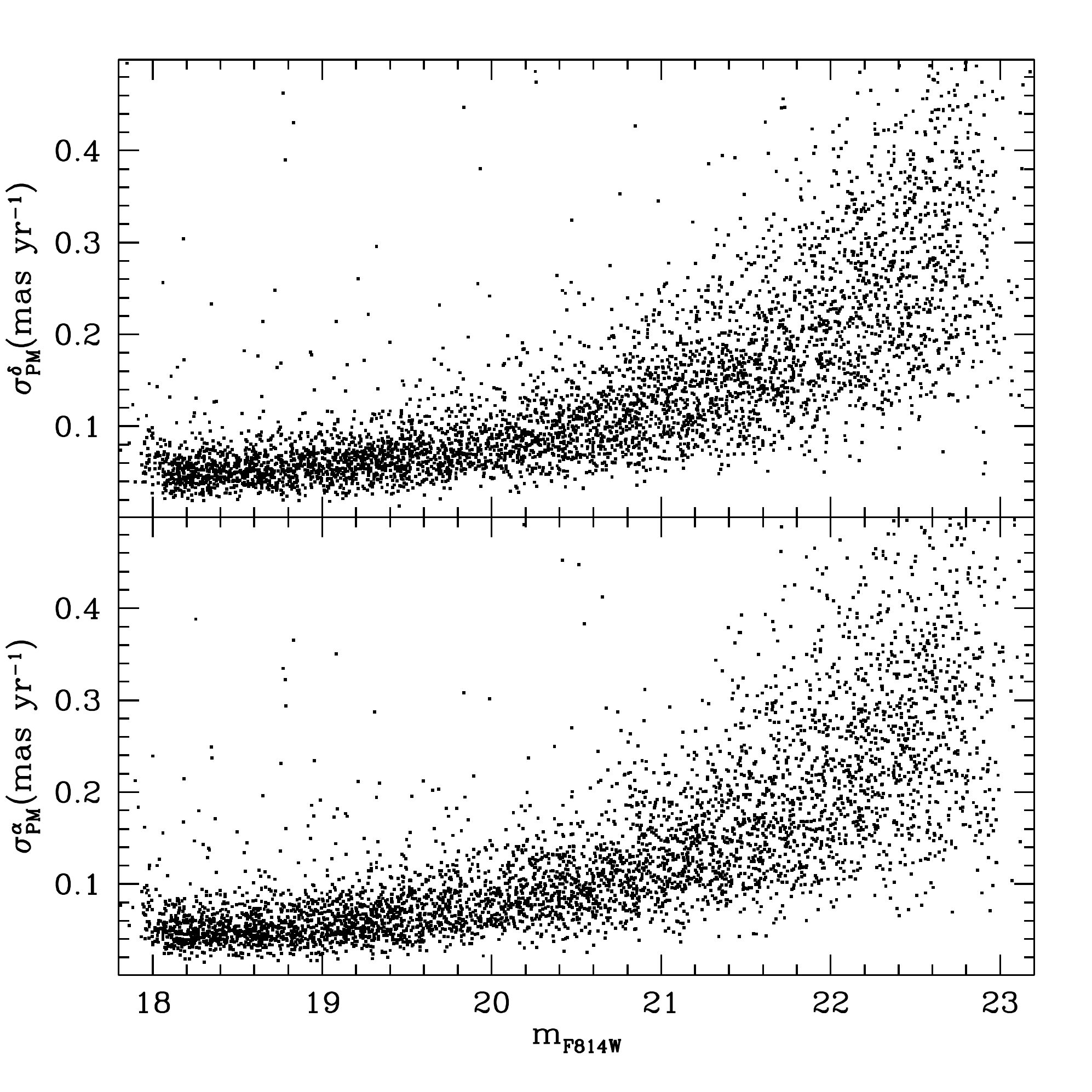}
\caption{Estimated uncertainties of the derived proper motions as a
  function of the $m_{\rm F814W}$ magnitude of the measured stars. The
  upper and the lower panels show, respectively, the uncertainties in
  the RA and in the Dec directions. For stars with $\rm m_{\rm
    F814W}\lesssim 21$ the typical error is smaller than $\rm 0.07
  \ mas \ yr^{-1}$. }
\label{errmoti}
\end{figure}

To study the PMs of M71, we used the high resolution datasets. These
are separated by a temporal baseline of 7.274 years and {because of
their different orientation, pointing and magnitude limit, the PM analysis could be performed only on the $\sim5000$ stars in common, located in the overlapping FOV} (see Figure~\ref{mappafov}) and
having magnitudes $18<m_{\rm F814W}<24$ (corresponding to magnitudes $19<m_{\rm F606W}<25$).
 We adopted the procedure described in \citet[][see also \citealp{dalessandro13,
    bellini14, massari15}]{massari13}. {Briefly, we used six parameters linear transformations\footnote{{To do this we applied six parameters linear transformations using CataXcorr, a code developed by P. Montegriffo at INAF- Osservatorio Astronomico di Bologna. This package is available at http://davide2.bo.astro.it/?paolo/Main/CataPack.html, and has been successfully used in a large number of papers by our group in the past 10 years. }} to report the coordinates of the stars in each exposure to the distortion-free reference catalog of
\citet{sarajedini07}. Since we are interested in the stellar PMs relative to the cluster frame, these transformations have been determined by using a sample of $\sim6600$ stars that, in the reference catalog, are likely cluster members on the basis of their CMD position (i.e. stars located along the main sequence). Moreover, the transformations have been determined independently on each detector chip in order to maximize the accuracy. At the end of the procedure, for each of the $\sim5000$ stars we have up to ten position measurements in the first epoch catalog and up to nineteen in the second epoch catalog.}  To determine the relative PMs, we computed the mean X and Y positions of each star in
each epoch, adopting a $3\sigma$ clipping algorithm. The star PMs are
thus the difference between the mean X,Y positions evaluated in the
two epochs, divided by $\rm \Delta t=7.274$ years. The resulting PMs
are in units of pixels $\rm years^{-1}$. Since the master frame is
already oriented according to the equatorial coordinate system, the
X-component of the PM corresponds to a projected PM along the
(negative) RA and the Y-component corresponds to a PM along the
Dec. Therefore, we converted our PMs in units of mas $\rm years^{-1}$
by multiplying the previous values for the ACS pixel scale ($\rm
0.05\arcsec pixel^{-1}$), and we named $\mu_\alpha \cos(\delta)$ and $\mu_\delta$
the PMs along the RA and Dec directions, respectively.  To maximize
the quality of our results, we built a final PM catalog by taking
into account only stars for which at least three position measurements
are available in each epoch. At the end of the procedure we counted
4938 stars with measured PMs. The errors in the position of the stars
in each epoch ($\rm \sigma_{1,2}^{\alpha,\delta}$) have been
calculated as the standard deviation of the measured positions around
the mean value. Then the errors in each component of the PM have been
assumed as the sum in quadrature between the error in the first and
second epoch: $\sigma_{\rm PM}^{\alpha,\delta} = \sqrt{
  (\sigma_1^{\alpha,\delta})^2 + (\sigma_2^{\alpha,\delta})^2 }/
\Delta t$. The errors as a function of the star magnitudes are shown
in Figure \ref{errmoti}. For both the PM directions, the typical
uncertainty for stars with $\rm m_{F814W}<21$ is less of $\rm \sim
0.07 \ mas \ yr^{-1}$, demonstrating the good quality of our
measurements.  {Following \citet{bellini14}, we also verified that our PM measurements are not affected by chromatic effects, i.e., there is no dependence of $\mu_\alpha \cos(\delta)$ and $\mu_\delta$ on the (F606W-F814W) color.}
Finally, our PM measurements are not even affected by positional
effects, i.e., there is no dependence of the derived PMs on the
instrumental (X,Y) positions.

\begin{figure}[t]
\centering
\includegraphics[width=8.5cm]{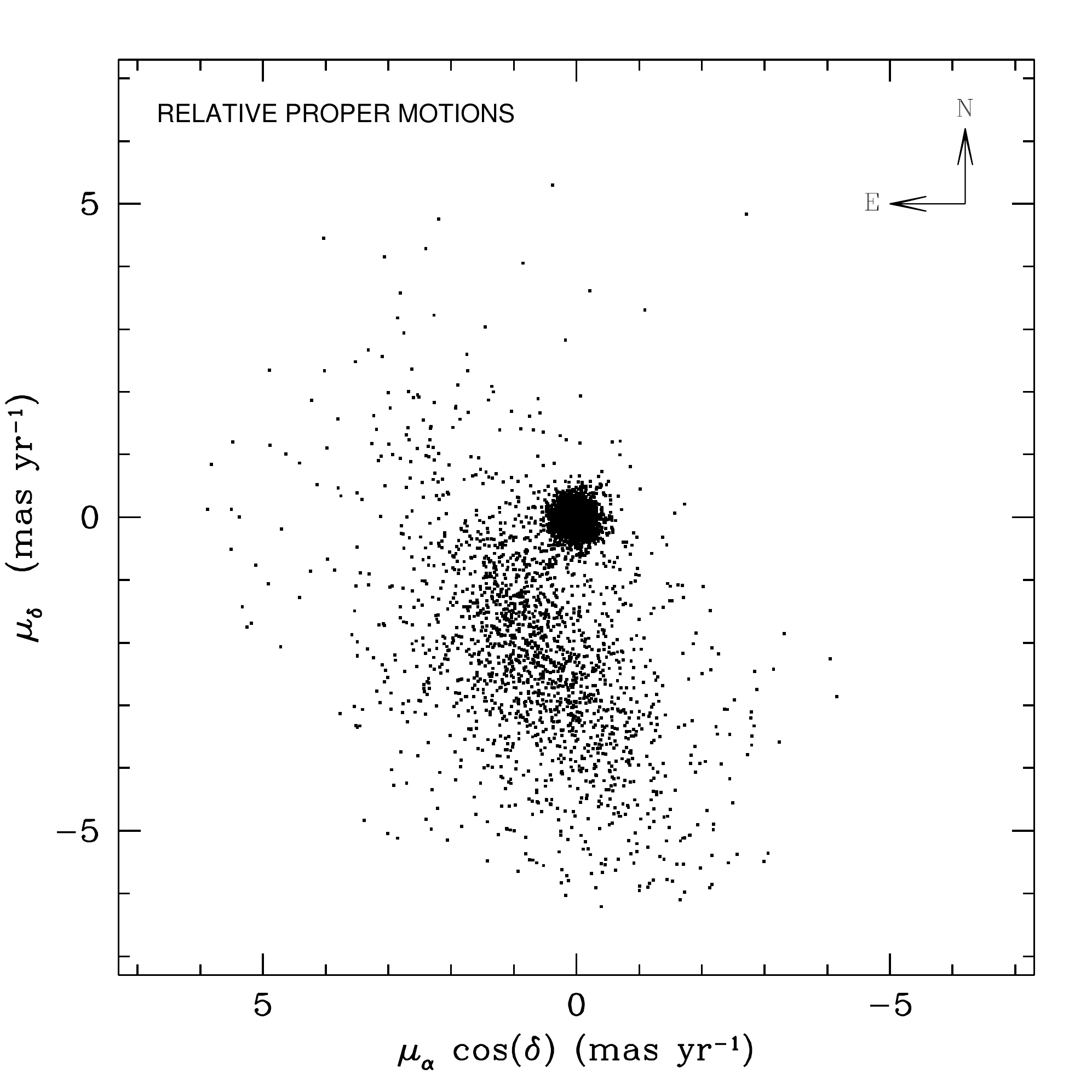}
\caption{VPD of the relative PMs. The clump in the (0,0) $\rm mas
  \ yr^{-1}$ position is dominated by the cluster population. The
  elongated region beyond this clump is instead due to contaminating
  stars, mostly from the Galactic disk.}
\label{motirel}
\end{figure}

\begin{figure}[!h]
\centering
\includegraphics[width=8.5cm]{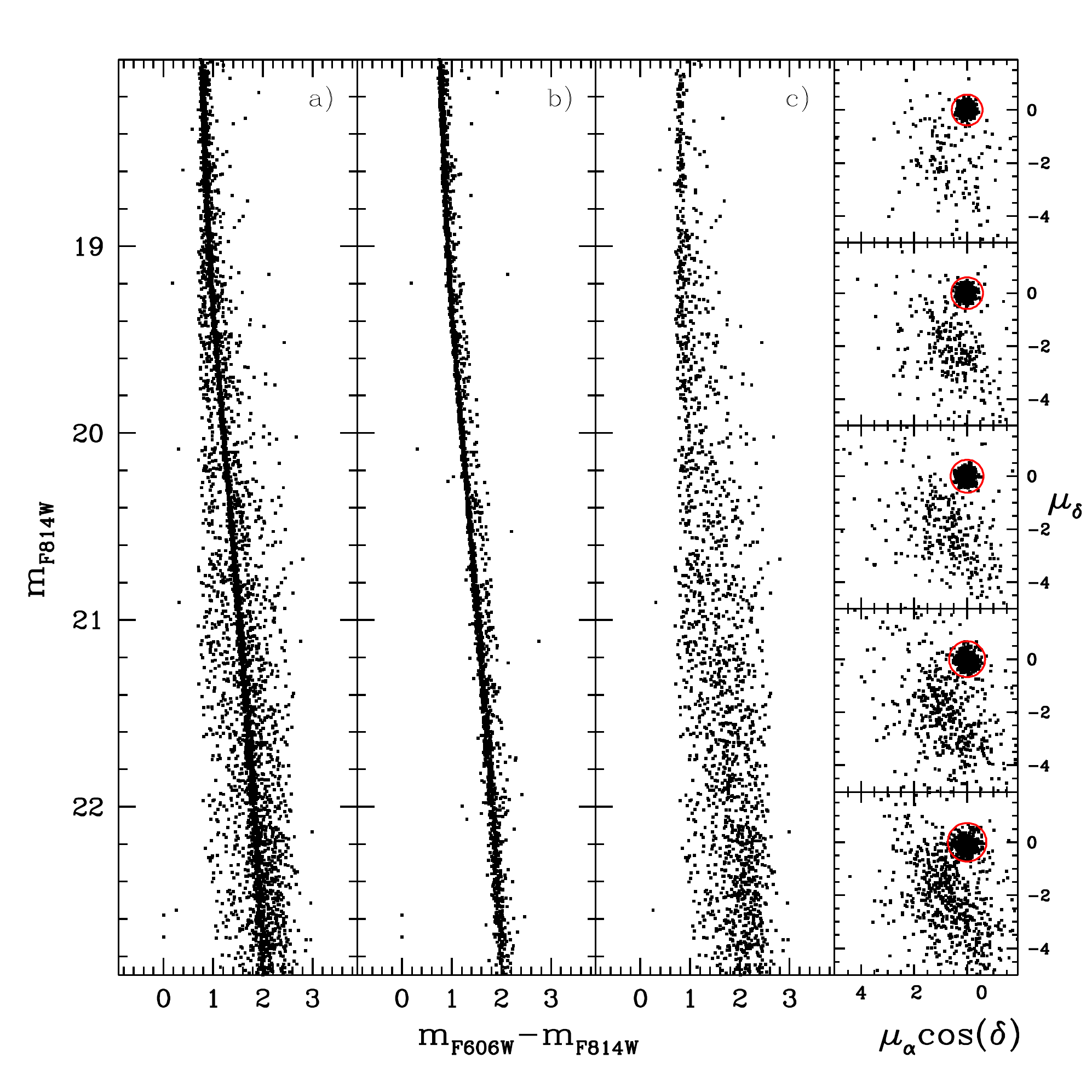}
\caption{{\it Panel a:} Optical CMD of all the stars in common
  between the two observation epochs. {\it Panel b:} Decontaminated
  CMD obtained by using only the likely cluster members selected from
  the VPDs shown in the rightmost column. As can be seen, a sharper
  and more delineated main sequence and the binary sequence are now
  appreciable. {\it Panel c:} CMD made of all the contaminating
  objects, selected from the VPDs as those with PMs not compatible
  with the that of the GC. The Galactic sequence can be appreciate
  from this plot. {\it Rightmost column:} VPDs of the measured stars
  divided in bins of magnitudes. The solid red circles contain all the
  objects selected as likely cluster members.}
\label{cmdmoti}
\end{figure}

\begin{figure*}[t] 
\centering
\includegraphics[width=8.5cm,angle=-90]{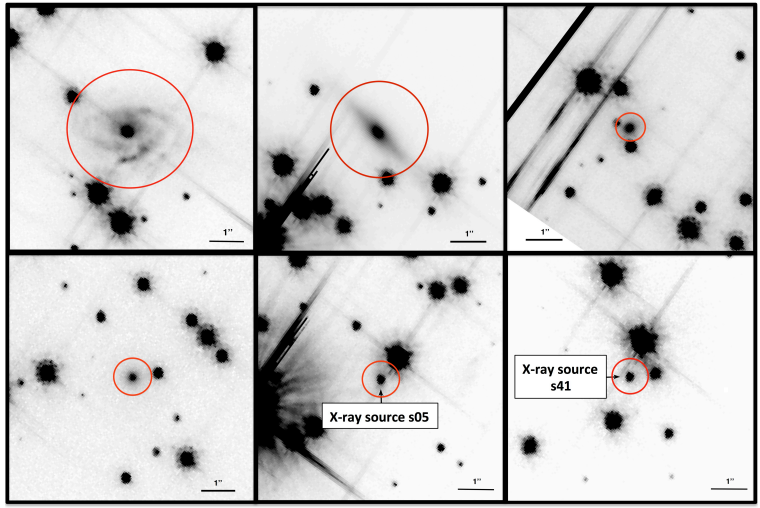}
\caption{Finding charts of the six selected extra-Galactic objects
  used to determine the absolute reference frame zero point. The
  charts are taken from an ACS image acquired in the F814W filter. The
  three upper panels and the bottom-left panel show the four galaxies
  found through the visual inspection of the images.  The central and
  right bottom panels show the optical counterparts to the X-ray
  sources s05 and s41, classified as AGNs. The point-like structure of
  these sources allowed a precise determination of the centroid
  position and of their PM.}
   \label{charts}
\end{figure*}

In Figure \ref{motirel} we show the PM distribution in the vector
points diagram (VPD). As can be seen, the VPD is dominated by two
prominent features: the clump in the center with zero relative PM is,
by definition, dominated by the cluster population, while the
elongated sparse distribution of points extending beyond this clump is
dominated by contaminating field stars, mostly from the Galactic
disk. {At first inspection of the VPD we can see that only $\lesssim
60\%$ of the $\sim5000$ analyzed stars are likely cluster members}. A high
percentage of field contamination is indeed expected in the case of
M71, since it has a quite low stellar density and is located in a
crowded field at low Galactic latitudes. By selecting in the VPD the
likely cluster members (i.e. the stars with relative PM around 0 both
in RA and in Dec) we find that the mean motion is $\rm 0.01 \ mas
\ yr^{-1}$ with a standard deviation of $\rm 0.1 \ mas \ yr^{-1}$ both
in $\alpha$ and in $\delta$, thus, as expected, consistent with zero.

The effect of decontaminating the CMD from field stars is shown in
Figure \ref{cmdmoti}, where we have separated the objects with PM$\lesssim \rm
0.6 \ mas \ yr^{-1}$ (likely cluster members), from those with larger
PMs.  The selection of stars in the VPD is shown, per bins of one
magnitude, in the left-hand column of the figure, with the objects
having PM$\lesssim \rm 0.6 \ mas \ yr^{-1}$ encircled in red. The effect on
the CMD is shown in the other three columns: from left to right, the
observed CMD, the CMD of cluster members only, and the CMD of field
stars.  In the latter, it is well appreciable the main sequence of the
Galactic field.  Instead, the decontaminated CMD clearly shows a sharp
and well defined main sequence, also revealing the binary sequence.
The few stars on the blue side of the main sequence could be cluster
exotic objects, such as cataclysmic variables, X-ray binaries or
millisecond pulsars \citep[e.g.][]{ferraro01,pallanca10,cohn10,cadelano15},
where a main sequence companion star is heated by a compact object.  Nonetheless, we cannot completely rule out the
possibility that some of these stars are field objects with PMs
compatible with those of the cluster members.

\section{ABSOLUTE PROPER MOTIONS}

\label{abspm}
\begin{figure}[t]
\centering
\includegraphics[width=8.5cm]{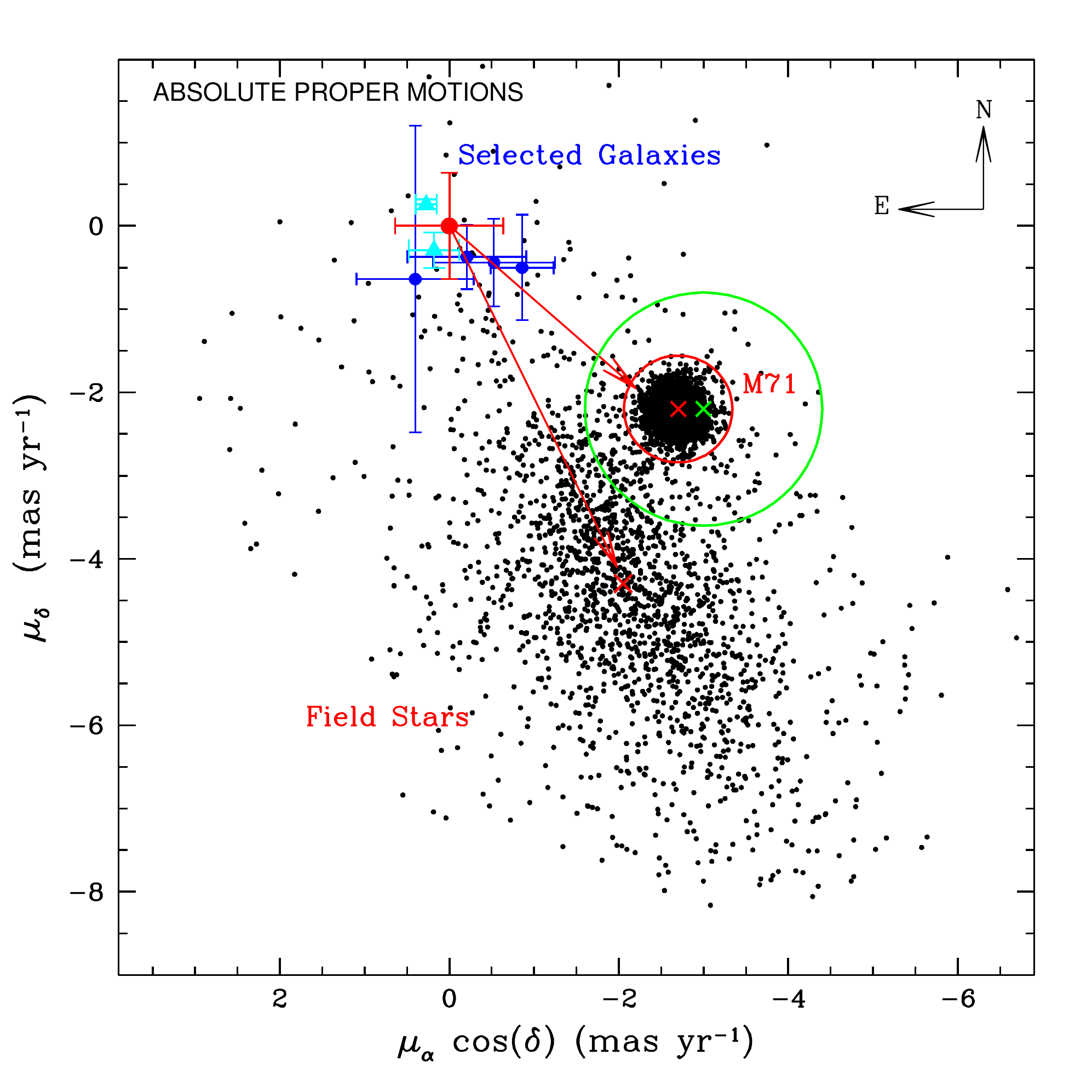}
\caption{VPD of absolute PMs. The extra-Galactic objects used to
  determine the absolute reference frame zero point are marked as blue
  dots (galaxies) and cyan triangles (AGNs), and their mean motion
  is marked a large red dot.  The red arrows indicate the absolute
  PM vectors of M71 and of the Galactic field population.  The
  absolute PM value of M71 estimated in this work is marked with a red
  cross and the red circle represents its $\sim1 \sigma$ confidence
  region. The values previously determined in the literature
  \citep{geffert00} are marked with a green cross and a green
  circle. The red cross centered on the elongated structure is centered on the
  field mean PM.}
  \label{motiass}
\end{figure}

To transform the relative PMs into absolute ones, we used
background galaxies as reference, since they have negligible PMs due
to their large distances. This method has been successfully used in
several previous works \citep[e.g.][]{dinescu99, bellini10,
  massari13}. Unfortunately, the NASA Extragalactic Database report no
sources in the FOV used for the PM estimate. Thus, we carefully
inspected our images in order to search for diffuse galaxy-like
objects. We found four galaxies with central point-like structure and
relative high brightness, which allowed us to precisely determine
their centroid position.  Although many other galaxies are present in
the FOV, they have no point-like structure or are too faint to allow
the determination of a reasonable PM value. Moreover, as part of a
project aimed at searching for optical counterparts to X-ray sources,
we identified two promising  active galactic nuclei (AGN) candidates. Two Chandra X-ray sources,
named s05 and s41 in \citet{elsner08}, have high energy and optical
properties that can be attributed either to AGNs or to cataclysmic
variables \citep[see][for more details]{huang10}. In order to
distinguish between these two possibilities, we analyzed their PMs. We
reported our relative PM reference frame to the absolute cluster PM
($\rm \mu_{\alpha}\cos \delta, \mu_{\delta}=-3.0\pm1.4, -2.2\pm1.4
\ mas \ yr^{-1}$) previously determined by \citet{geffert00} and found
that these two sources have an absolute PM significantly different
from the cluster motion and compatible with zero. We therefore
conclude that these two objects are likely background AGNs\footnote{Of
  course, these sources could be foreground cataclysmic variables with
  PMs almost perfectly aligned with our line of sight, but this
  possibility seems to be quite unlikely.} and add them to the list of
objects used to determine our reference absolute zero point.  The six
selected objects are located very close to each other in the VPD, as
expected for extragalactic objects, and their finding charts are shown
in Figure \ref{charts}. We defined the absolute zero point as the
weighted mean of their relative PMs and assumed as error the
uncertainty on the calculated mean. By anchoring this mean position to
the (0,0) mas $\rm yr^{-1}$ value, we find that the absolute PM of M71
is:
\begin{equation}
\rm \left(\mu_{\alpha}\cos \delta, \mu_{\delta}\right)=(-2.7\pm0.5,
-2.2\pm0.4) \ mas \ yr^{-1}.
\end{equation}
This value is in good agreement with (but more accurate than) the
previous determination \citep{geffert00}, and it remains unchanged
within the errors even if the two candidate AGNs are excluded from the
analysis: in that case we get: $\rm \left(\mu_{\alpha}\cos \delta,
\mu_{\delta}\right)=(-2.4\pm0.6, -1.9\pm0.1) \ mas \ yr^{-1}$, still
in agreement with the previous results.  The VPD in the absolute frame
is plotted in Figure \ref{motiass}, with the red and green crosses and
circles marking, respectively, the absolute PM and its uncertainty as
determined in this study and as quoted in \citet{geffert00}.

Since every absolute PM measurement is strictly dependent on the
accuracy of the absolute reference frame, we need to verify the
possible presence of systematic errors in its determination. One of
the possible source of systematic errors is the rotation of the GC on
the plane of the sky. Indeed, since we used only cluster stars to
define the relative reference frame, if the GC is rotating, then our
frame will be rotating too. This would introduce an artificial
rotation to background and foreground objects around the cluster
center. {To quantify this possible effect we followed the procedure described in \citet{massari13}. We selected a sample of field stars as those that in the VPD of Figure~\ref{motirel} have relative PMs larger than $\rm 0.8 \ mas \ yr^{-1}$. Then we decomposed their PM vectors into a 
radial and tangential component with respect to the cluster center. If the GC is rotating,
we would expect to find a clear dependence of the PM tangential component on the distance from the cluster center. Such a dependence is however excluded by our results, thus that the internal regions of M71 are not rotating, in agreement with the recent findings by \citet{kimmig15}}.

{We also compared} the field star motion to that expected
from theoretical Galactic models in the analyzed FOV. To evaluate the
field mean motion, we followed the procedure described in
\citet{anderson10}. First, we excluded the stars within $\rm 0.8 \ mas
\ yr^{-1}$ from the cluster mean motion. Then we iteratively removed
field stars in a symmetric position with respect to the GC exclusion
region and evaluated the weighted mean motion by applying a $3\sigma$
algorithm. We found $\rm \left(\mu_{\alpha}\cos \delta,
\mu_{\delta}\right)=(-2.0\pm0.2, -4.3\pm0.2) \ mas \ yr^{-1}$. We
compared these values with those predicted for the same region of the
sky in the Besan\c{c}on Galactic model \citep{robin03}, simulating a
sample of $\sim2000$ artificial stars distant up to 15 kpc from the
Galactic center, in a FOV centered on M71, covering a solid angle of
$\sim11\arcmin$, and having V magnitudes ranging 12 from to 25. The
predicted field mean motion is $\rm \left(\mu_{\alpha}\cos \delta,
\mu_{\delta}\right)=(-2.4, -4.7) \ mas \ yr^{-1}$, in good agreement
with our results.


\subsection{The cluster orbit}
\label{orbitaround}
\begin{figure*}[t]
\centering
\includegraphics[width=8cm]{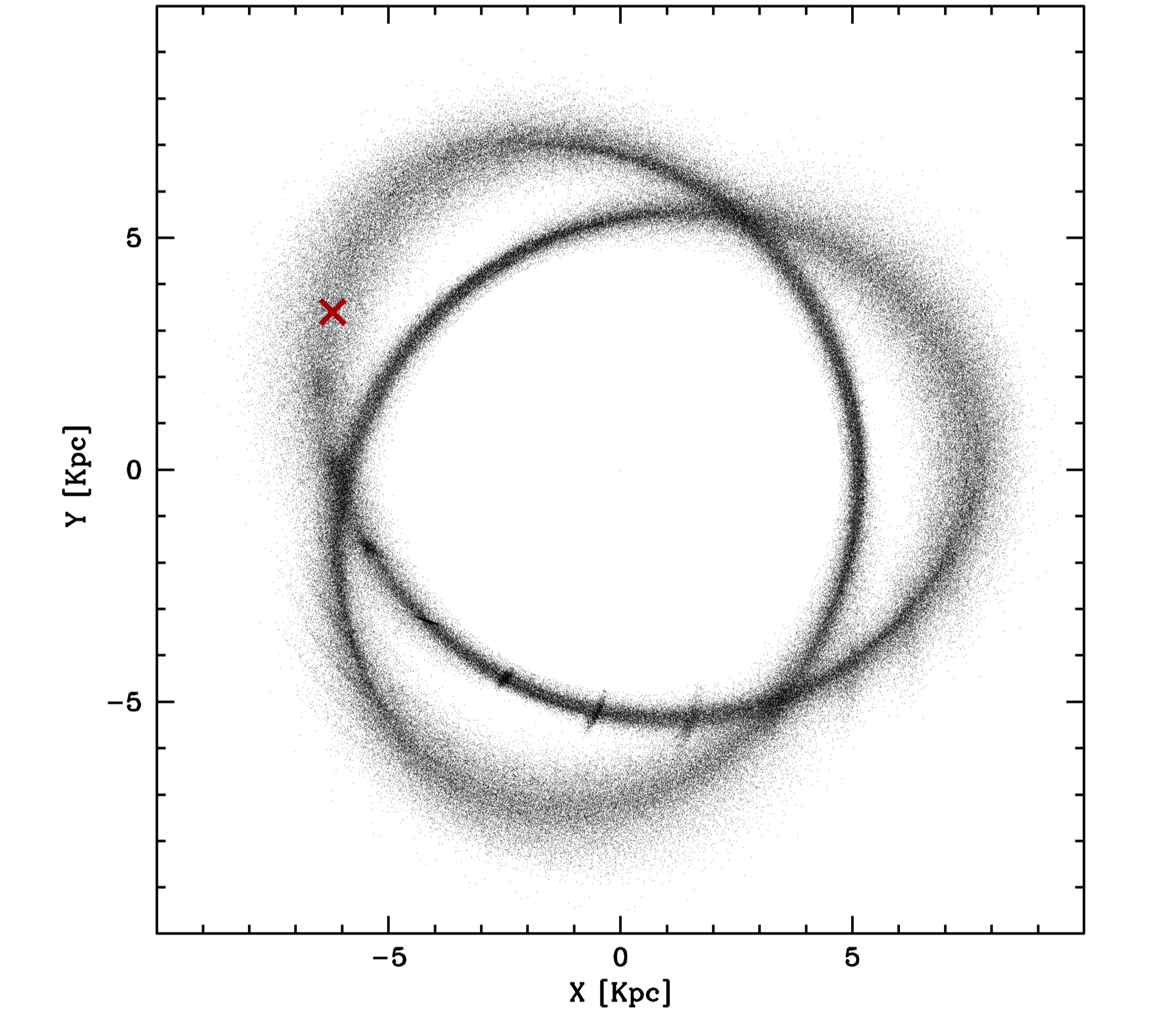}
\includegraphics[width=8cm]{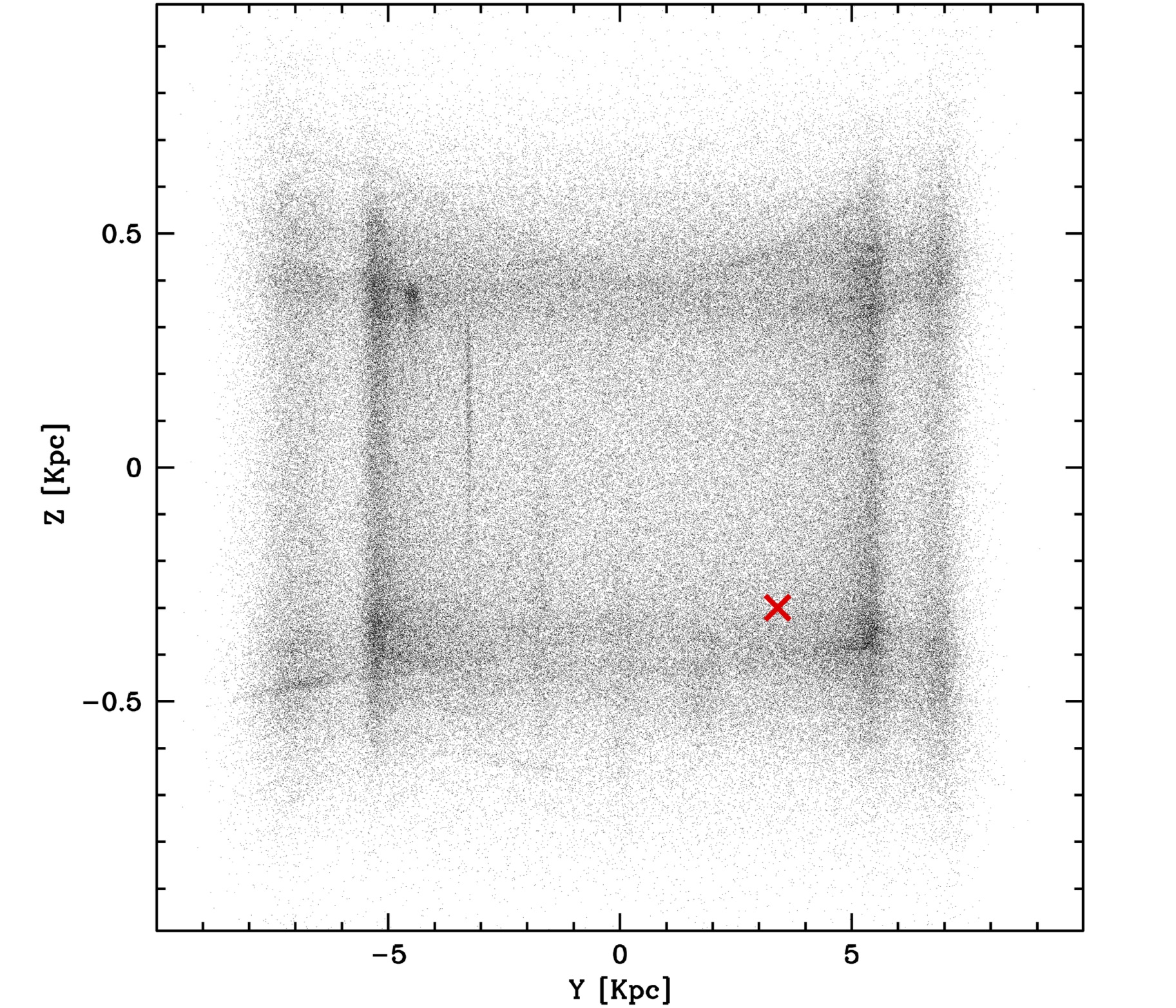}
\caption{{\it Left:} Simulated positions occupied by M71 during the
  last 3 Gyr along its orbit in the equatorial Galactic plane. Each
  point represents the position of one (out of 1000) cluster in one of
  the 32000 snapshots obtained during the numerical integration. The
  red cross marks che current cluster position. {\it Right:} Same as
  in the left panel, but for the orbit in the meridional Galactic
  plane.}
  \label{orbit}
\end{figure*}

The GC absolute PM, combined with the radial velocity
$v_r=-23.1\pm0.3$ km s$^{-1}$ from \citet{kimmig15}, can be used to
determine 3D space velocity of the cluster in a Cartesian
Galactocentric rest frame. Using the formalism described in
\citet{johnson87}, assuming the Local Standard of Rest {velocity equal
to $\rm 256 \ km \ s^{-1}$ \citep{reid09}} and using the value of the Sun velocity
with respect to it from \citet{schonrich10}, {we obtained $\rm
\left(v_x,v_y,v_z \right)=\left(52\pm10,204\pm6,31\pm12\right) \ km
\ s^{-1}$}, where the major source of uncertainty is the GC absolute PM
error.
We then used the 3D velocity of the cluster and its current
Galactocentric position\footnote{We adopted the Galactic coordinates
  quoted in \citet{harris96} and the convention in which the X axis
  points opposite to the Sun, i.e., the Sun position is $(-8.4,0,0)$, {where the distance of the Sun from the Galactic center is from \citet{reid09}}.}
$\rm \left(x,y,z \right)=\left(-6.2\pm0.6,3.4\pm0.3,-0.32\pm0.03
\right) \ kpc$, to reconstruct its orbit in the axisymmetric potential
discussed in \citet{allen91}, which has been extensively used to study
the kinematics of Galactic stellar systems
\citep[e.g.][]{ortolani11,moreno14,massari15}.  The orbit was
time-integrated backwards, starting from the current conditions and
using a second-order leapfrog integrator \citep[e.g.][]{hockney88}
with a small time step of $\sim100$ kyr. Since the adopted Galactic
potential is static, we choose to back-integrate the orbit only for 3
Gyr, since longer backward integrations become uncertain due to their
dependence on the Galactic potential variations as a function of
time. This numerical integration required about 32000 steps and
reproduced $\sim20$ complete cluster orbits. The errors on the
conservation of the energy and the Z-component of the angular momentum
never exceeded one part over $10^9$ and $10^{16}$, respectively. We
generated a set of 1000 clusters starting from the phase-space initial
conditions normally distributed within the uncertainties. For all of
these clusters we repeated the backward time
integration. Figure~\ref{orbit} shows the resulting cluster orbits in
the equatorial and meridional Galactocentric plane. As can be seen,
the cluster has a low-latitude disk-like orbit within the Galactic
disk. Indeed in the equatorial plane it reaches a maximum distance of
$\sim8$ kpc from the Galactic center and a minimum distance of $\sim5$
kpc. Thus, it orbits around the assumed spheroidal bulge, never
crossing it. Moreover, it persists on a low-latitude orbit, with a
typical height from the Galactic plane of about $\pm0.4$ kpc, thus
again confined within the disk. The estimated orbits indicate that, at
least during the last 3 Gyr, M71 tightly interacted with the inner
Galactic disk. {With respect to the large majority of Galactic GCs, which are on large orbits across the (low-density) halo, these interactions likely induced heavy mass-loss  \citep{vesperini97} in M71, thus supporting the possibility that it lost a significant fraction of its initial mass,} as already suggested by its flat mass function
\citep{demarchi07}. Moreover, such a heavy mass-loss could finally
explain why M71 harbors a large population of X-ray sources, in spite
of its present low mass \citep{elsner08}.

\section{GRAVITATIONAL CENTER, STRUCTURAL PARAMETERS AND INITIAL MASS}
\label{clust}
In this section we present the determination of the gravitational
center and of the new structural parameters of M71.
 
\subsection{Gravitational center}
\label{center}
To avoid biases due to the strong differential reddening affecting the
system \citep[e.g.][]{schlegel98}, for the determination of the
cluster center of gravity $C_{\rm grav}$ we used the near-infrared
WIRCam catalog, which has the same level of completeness of the ACS
one in the magnitude range $14 < K_s < 16.8$. {$C_{\rm grav}$ has been determined following an iterative procedure that, starting from a first-guess center, selects a sample of stars within a circle of radius r and re-determine the center as the average of the star coordinates ($\alpha$ and $\delta$). The procedure stops when convergence is reached, i.e., when the newly-determined center coincides with the previous ones within an adopted tolerance limit \citep[][see also \citealt{montegriffo95,lanzoni07}]{lanzoni10}.} 
For M71,
which is a relative loose GC
\citep{harris96}, we repeated the procedure eighteen times, using
different values of $r$ and selecting stars in different magnitude
ranges, chosen as a compromise between having high enough statistics
and avoiding spurious effects due to incompleteness and saturation.
In particular, the radius $r$ has been chosen in the range
$140\arcsec-160\arcsec$ with a step of $10\arcsec$, thus guaranteeing
that it is always larger than the literature core radius
$r_c=37.8\arcsec$ \citep{harris96}. For each radius $r$, we have
explored six magnitude ranges, from $K_s>14$ (in order to exclude stars
close to the saturation limit), down to $K_s=16.3-16.8$, in steps of 0.1
magnitudes. As first-guess center we used that quoted by
\citet{goldsbury10}.  The final value adopted as $C_{\rm grav}$ is the
mean of the different values of RA and Dec obtained in the eighteen
explorations, and its uncertainty is their standard deviation. We
found $\rm RA=19^{h}53^{m}46.106^{s}$ and
$Dec=+18^{\circ}46\arcmin43.38\arcsec$, with an uncertainty of about
$1.7\arcsec$.  The newly determined center of M71 is $\sim 5.7\arcsec$
west and $\sim 0.3.\arcsec$ north from the one measured from optical
ACS data by \citet{goldsbury10}. Such a discrepancy is likely
ascribable to an effect of differential reddening impacting the
optical determination. 

\subsection{Stellar density profile}
\label{density}
\begin{figure}[h]
\centering
\includegraphics[width=8.5cm]{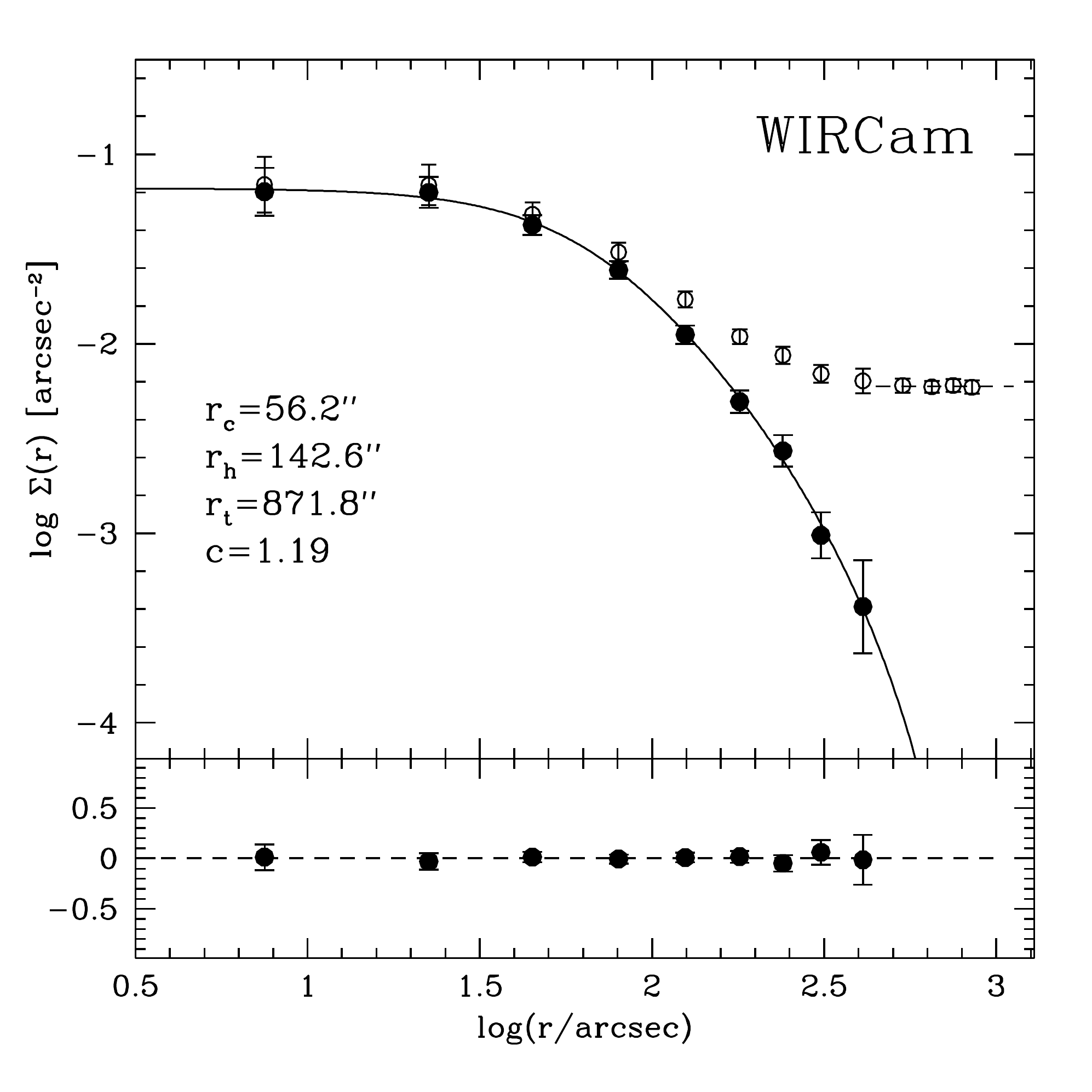}
\caption{Observed density profile of M71 obtained from the WIRCam
  near-infrared dataset (open circles). The dashed line marks the
  density value of the Galactic field background, obtained by
  averaging the four outermost points. The black filled circles are
  the density values obtained after background subtraction (see
  Section \ref{density}). The best-fit King model (solid line) is
  overplotted to the observations and the residuals of the fit are
  reported in the bottom panel. The best-fit structural parameters are
  also labelled in the figure.}
\label{profwircam}
\end{figure}

\begin{figure*}[t]
\centering
\includegraphics[width=8.1cm]{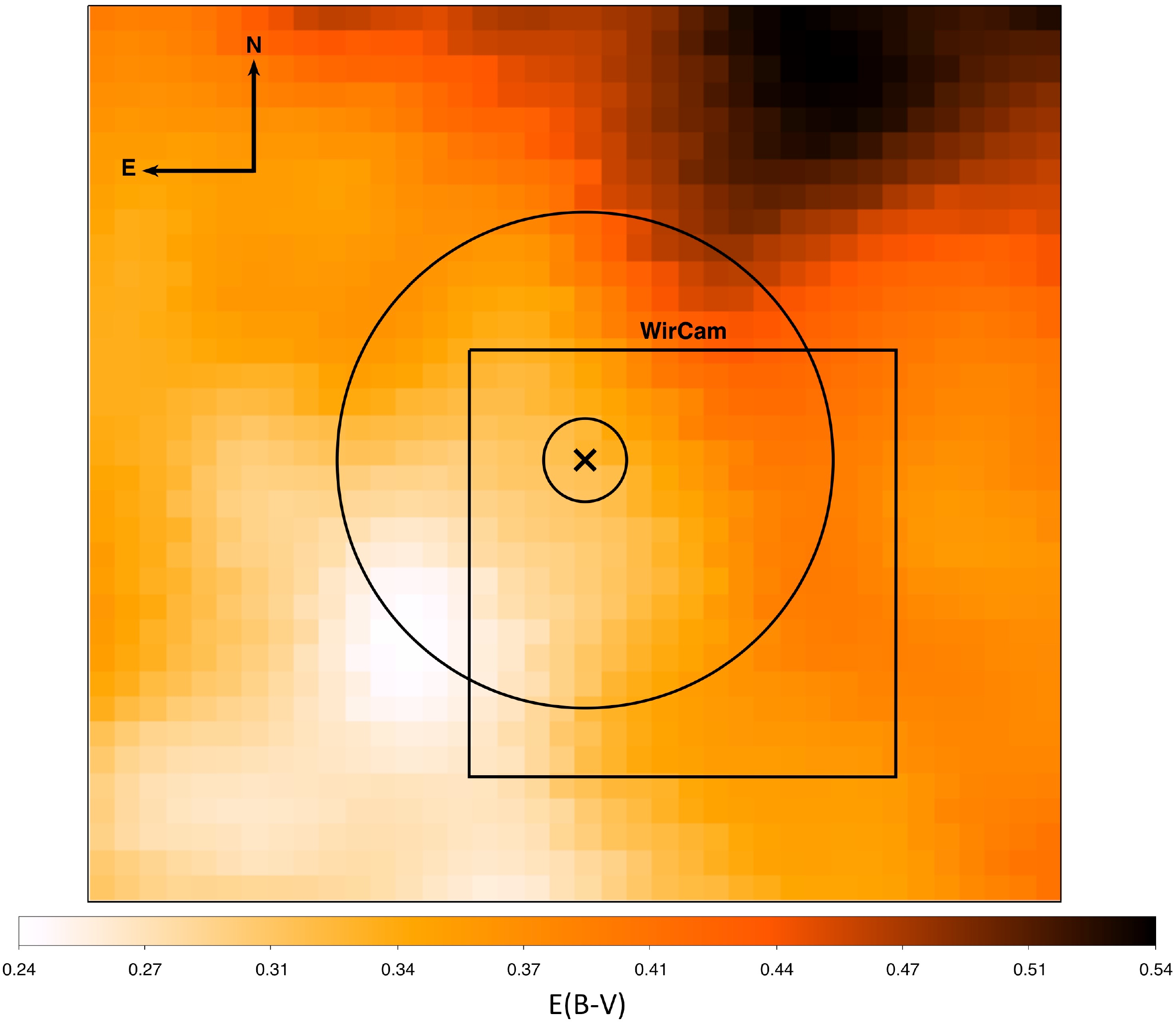}
\includegraphics[width=8.2cm]{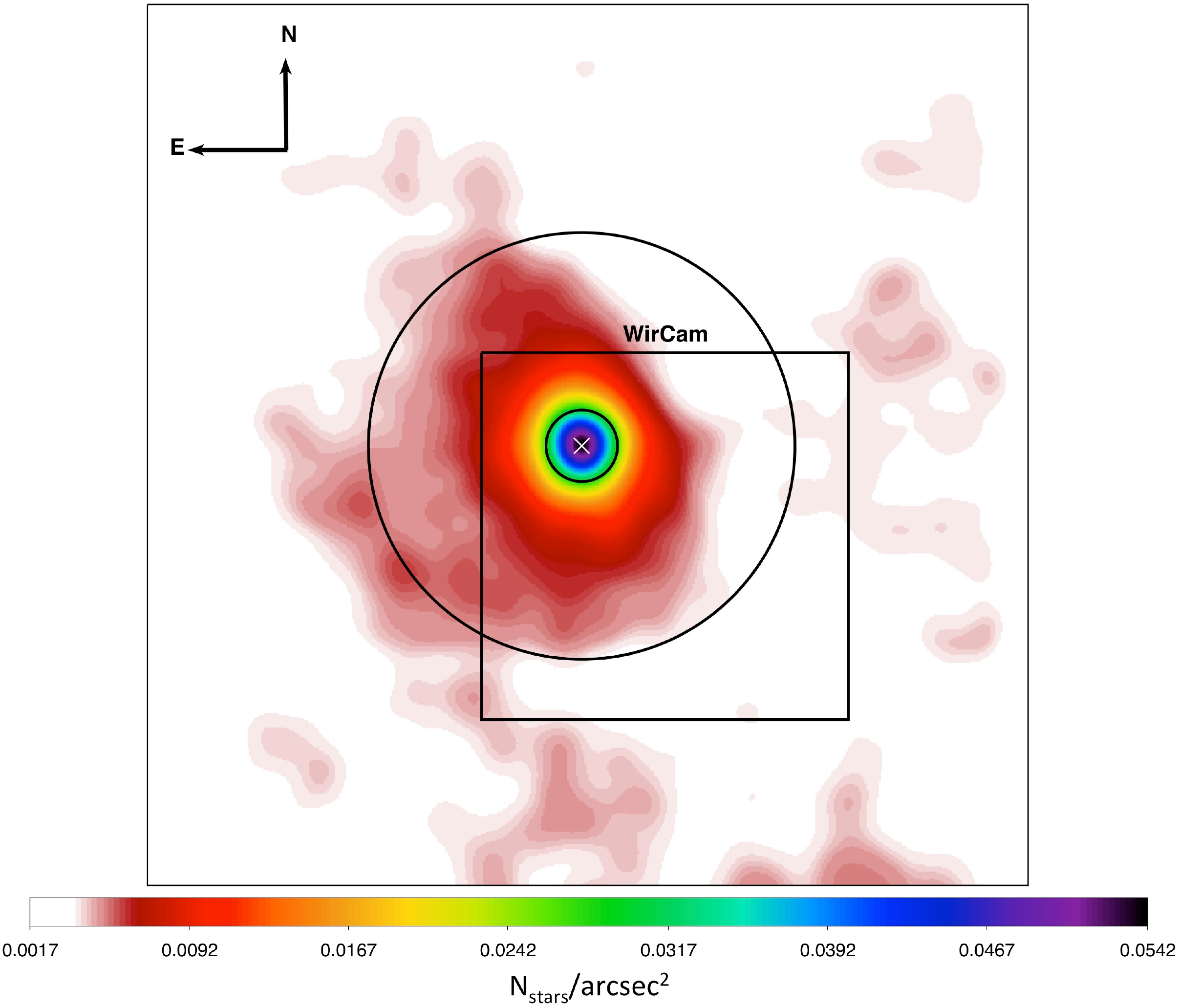}
\caption{{\it Left Panel:} Reddening map (from \citealp{schlegel98})
  of the $1500\arcsec\times 1500\arcsec$ region centered on M71. The
  small and large circles mark, respectively, the half-mass and
  truncation radii of the cluster. The square marks the FOV of the
  WIRCam dataset. The presence of severe differential reddening across
  the system, especially in the north-west sector, is apparent.  {\it
    Right Panel:} 2D stellar density map in the same region of the left panel, obtained by counting all the stars with
  $13<g'<19$ in the optical (MegaCam) dataset. The comparison with the
  reddening map on the left makes the effect of extinction well
  visible: the significant decline in the stellar density observed in
  the external regions of the map (especially in the north-east
  sector) is clearly due to the presence of thick dust clouds.}
\label{mapred}
\end{figure*}

Since the surface brightness profile can suffer from strong biases and
fluctuations due to the presence of few bright stars \citep[see, e.g.,
  the case of M2 in][]{dalessandro09}, in order to re-evaluate the
structural parameters of M71 we used direct star counts.  The
determination of the stellar density profile (number of stars per unit
area, in a series of concentric annuli around $C_{\rm grav}$) has been
performed following the procedure fully described in
\citet{miocchi13}. Also in this case, in order to minimize the
differential reddening effect we used the near-infrared WIRCam data,
which covers distances out to $\sim 1000\arcsec$ from $C_{\rm grav}$
in the south-west portion of the cluster (see Fig. \ref{mapground}).
To build the density profile we considered 13 concentric annuli around
$C_{\rm grav}$, each one divided into four sub-sectors. We then
counted the number of stars with $14<K_s<16.5$ in each sub-sector and
divided it by the sub-sector area.  The projected stellar density in
each annulus is the mean of the values measured in each sub-sector and
the uncertainty has been estimated from the variance among the
sub-sectors. The stellar background has been estimated by averaging
the outermost values, where the profile flattens, and it has been
subtracted to the observed distribution to obtain decontaminated
density profiles.  The result is shown in Figure \ref{profwircam}.

The cluster structural parameters has been derived by fitting the
observed density profiles with a spherical, isotropic, single-mass
\citet{king66} model.\footnote{These models can be generated and
  freely downloaded from the Cosmic-Lab web site:
  \url{http://www.cosmic-lab.eu/Cosmic-Lab/Products.html}. The fitting
  procedure is fully described in \citet{miocchi13}}. The single-mass
approximation is justified by the fact that the magnitude range chosen
to build the profile includes cluster stars with negligible mass
differences.  The best-fit model results in a cluster with a King
dimensionless potential $W_0=5.55\pm 0.35$, a core radius
$r_c=56.2^{+4.5}_{-4.0}$ arcsec, a half-mass radius $r_h=
146.2^{+11.5}_{-10.0}$ arcsec, a truncation radius $r_t=
871.8^{+247}_{-164}$ arcsec and, thus, a concentration parameter,
defined as the logarithm of the truncation to the core radius,
$c=\log r_t/r_c=1.19$.

{There is a significant difference between these parameters and those quoted in the \citet{harris96} catalog, originally estimated by \citet{peterson97} from a surface brightness profile obtained from shallow optical images: rc = $37.8\arcsec$, $r_h = 100.2\arcsec$ and $r_t = 533.9\arcsec$ (the latter being derived from the quoted value of the concentration parameter: $c = 1.15$). To further investigate this discrepancy, we built the cluster surface brightness profile using a K-band 2MASS image, and we found that it is in agreement with the number density profile shown in Figure~\ref{profwircam}, thus further reinforcing the reliability of the newly-determined parameters. On the other hand, if we take into account only the brightest pixels of the K-band image, we find a surface brightness profile consistent with the literature one. This implies that the structural parameters quoted in the literature (which are determined from the light of the most luminous giants only) are not representative of the overall cluster profile.}

The availability of a very wide ($\sim1^{\circ}\times1^{\circ}$)
sample at optical wavelengths (the MegaCam dataset) with an analogous
level of completeness (comparable to the ACS one for $13<g'<19$)
allowed us to investigate how the derivation of the cluster stellar number 
density profile from optical observations can be affected by the
presence of large differential extinction.  Figure \ref{mapred}
compares the extinction map and the 2D density map of the
$1500\arcsec\times 1500\arcsec$ region of the sky centered on M71. The
former is obtained from \citet{schlegel98} and shows that the color
excess $E(B-V)$ varies from $\sim 0.24$ to $\sim 0.54$, with several ``spots'' and a clear gradient across
the field. The density map in the right-hand panel shows the number of
stars with $13<g'<19$, per unit area, detected in the MegaCam
sample. As expected, at large scales it reveals a direct
correspondence with the extiction map: in particular, the stellar
density manifestly drops in the north-west sector, where the color
excess is the highest, while the opposite is true in the south-east
part of the cluster. Obviously, this is expected to significantly
impact the density profile obtained from star counts in the optical
bands.

\begin{figure}[!h]
\centering
\includegraphics[width=8.5cm]{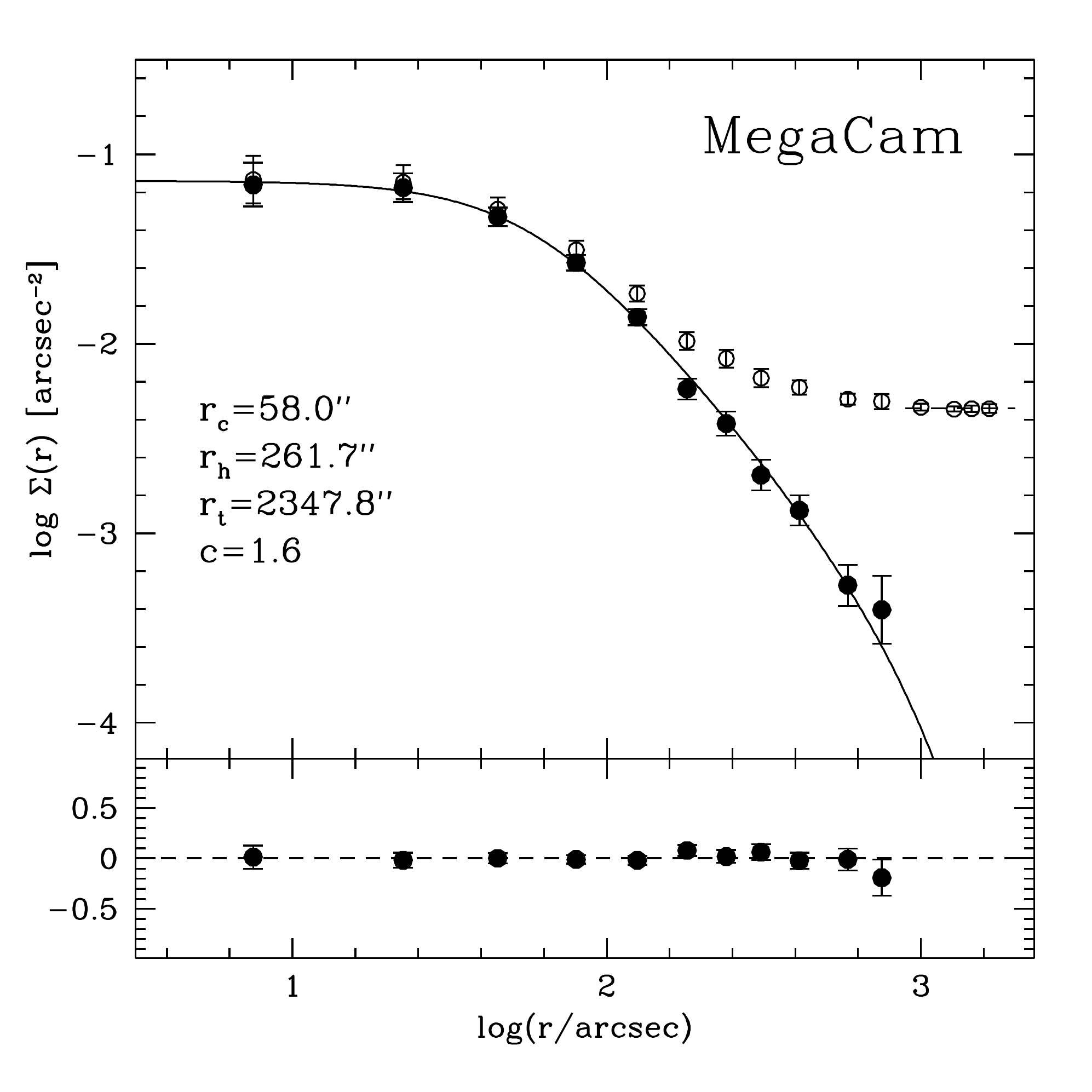}
\caption{Density profile of M71 obtained from the optical (MegaCam)
  dataset. The meaning of all symbols and lines is as in
  Fig. \ref{profwircam}. The best-fit King model parameters
  (especially the truncation radius and the concentration) are
  significantly different from those obtained from near-infrared
  observations, mainly because the Galactic background is
  underestimated at optical wavelengths, due to the large differential
  reddening affecting the external regions of the covered FOV (compare
  to Fig. \ref{mapred}).}
\label{profmega}
\end{figure}

To quantitavely test this effect, we determined the cluster density
profile by using the MegaCam (optical) data. The result is plotted in
Figure \ref{profmega} and shows that, indeed, the stuctural parameters
of the best-fit King model turn out to be very different from those
obtained from the near-infrared (almost reddening-unaffected) dataset
(compare with Fig. \ref{profwircam}). In particular, the concentration
parameter is much larger ($c=1.6$), as a consequence of a comparable
core radius ($r_c=58\arcsec$ \emph{versus} $56.2\arcsec$), but a more
than doubled truncation radius ($r_t=2347.8\arcsec$ \emph{versus}
$871.8\arcsec$).  Such a severe over-estimate of $r_t$ is due to the
high extinction affecting the external portions of the MegaCam sample,
where the Galactic field background is evaluated, and it clearly
demonstrates how important is to take differential reddening under
control for the determination of a cluster density profile. 

\subsection{Cluster initial mass}
\label{in_mass}
In Sect. \ref{orbitaround} we have argued that M71 likely lost a
significant fraction of its original mass, mostly due to environmental
effects. In this section, we attempt to estimate the total cluster
initial mass. Although many recipes can be used to this aim
\citep[e.g.][]{vesperini13}, we adopted the simple analytical approach
described in \citet{lamers05,lamers06}. It describes the way a cluster
loses its mass due to the effects of both stellar and dynamical
evolution (including processes such as interactions with the Galactic
tidal field and shocks due to encounters with giant molecular clouds
or spiral arms).  Although this method has been developed specifically
for open clusters, it can be used also in the case of M71, since its
current mass ($M=2.0^{+1.6}_{-0.9}\times10^4 M_\odot$; from
\citealp{kimmig15}) and orbit are consistent with those typical of
open clusters \citep[see also][for a similar implementation of this
  procedure]{dalessandro15}. The initial mass $M_{\rm ini}$ of the
cluster can be expressed as follows:
\begin{equation}
M_{\rm ini}\simeq \left[ \left( \frac{M}{M_\odot}\right)^\gamma +
  \frac{\gamma t}{t_0}\right]^{\frac{1}{\gamma}} \left[
  1-q_{ev}(t)\right]^{-1}, 
\end{equation}
where $M$ is the cluster current mass, $t=12\pm1$ Gyr is the cluster
age \citep{dicecco15}, $t_0$ is the dissolution time-scale parameter,
$\gamma$ is a dimensionless index and $q_{ev} (t)$ is a function
describing the mass-loss due to stellar evolution. The dissolution
time-scale parameter is a constant describing the mass-loss process,
which depends on the strength of the tidal field.  Small values of
$t_0$ are typically associated with encounters with molecular clouds
and spiral arms, while larger values are used to describe the effect
of the Galactic tidal field \citep[see][]{lamers05}.  Since M71 has an
orbit and a structure quite similar to those of open clusters, we
assumed $t_0$ in the same range of values ($2.3<t_0<4.7$ Myr)
constrained in \citet{lamers05}.  The parameter $\gamma$ depends on
the cluster initial density distribution and is usually constrained by
the value of the King dimensionless potential $W_0$. We adopted $\gamma = 0.62$,
corresponding to $W_0 = 5$, a typical value for an averagely concentrated cluster.
The function $q_{ev} (t)$, which describes the
mass-loss process due to stellar evolution, can be approximated by the
following analytical expression:
\begin{equation}
\log q_{ev}(t) = (\log t-a)^{b} +c, \ \ {\rm for} \ t > 12.5 \ {\rm
  Myr},
\end{equation}
where $a$, $b$ and $c$ are coefficients that depend on the cluster
metallicity.  The iron abundance ratio of M71 is [Fe/H]=$-0.73$
\citep{harris96}, which corresponds to $a=7.03$, $b=0.26$ and
$c=-1.80$ \citep{lamers05}.

\begin{figure}[t]
\centering
\includegraphics[width=8.5cm]{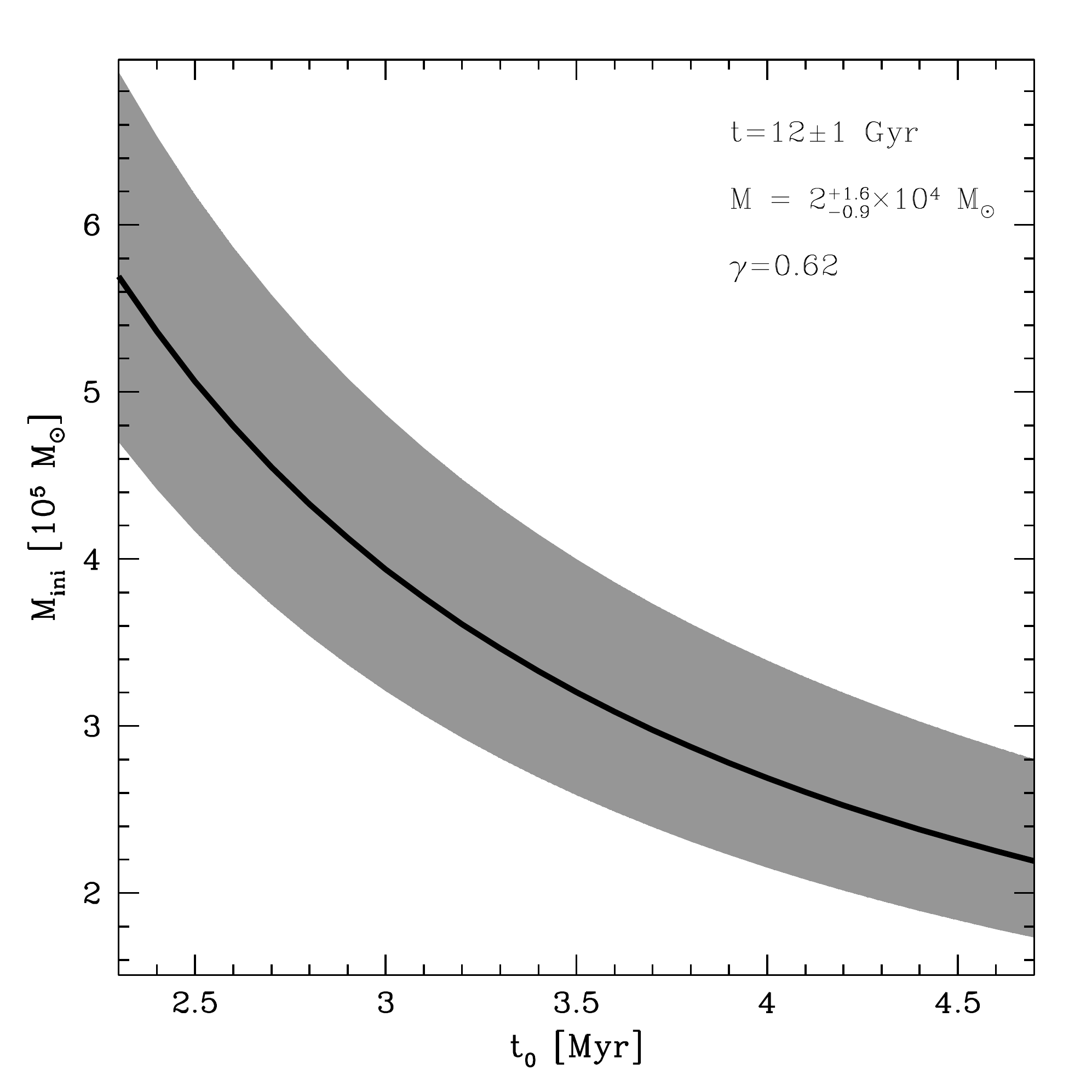}
\caption{Initial mass of M71 as a function of the dissolution
  time-scale parameter $t_0$, estimated as described in
  Sect. \ref{in_mass}. The black curve shows the values obtained by
  assuming a cluster current mass of $2\times10^4 \ M_\odot$ and an
  age of 12 Gyr. The gray shaded area includes all the values
  determined by considering the uncertainties in these quantities (see
  labels). }
  \label{mini}
\end{figure}

The resulting initial mass of the cluster is shown in Figure
\ref{mini} as a function of the explored range of values of $t_0$.  It
varies between 1.8 and $6.8\times10^5 M_\odot$, which are all values
typical of the mass of Galactic halo GCs, and is one order of
magnitude (or more) larger than the current mass.  Also considering
the largest possible value of $t_0$ ($\sim30$ Myr; see
\citealp{lamers05}), we find that the cluster initial mass is at least
twice its current value.  Clearly, this estimate is based on a
simplified approach and on parameters derived by the average behaviors
of open clusters, and different assumptions may lead to different
results. {However, it is interesting to note that, while such a high mass loss would be unlikely for a halo GC, it can be reasonable for a system moving along an orbit confined within the disk (see Sect.~\ref{orbitaround}).}

\section{SUMMARY AND CONCLUSIONS}
\label{conclu}
{By using two high-resolution ACS datasets separated by a temporal baseline of $\sim7$ years, we determined the relative PMs of $\sim5000$ individual stars in the direction of the low-mass GC M71, finding that only $\sim60\%$ of them have PMs consistent with being members of the cluster.} The identification of four galaxies and two AGNs within the
sampled FOV, allowed us to also constrain the absolute PM of M71. This
has been used to infer the orbit of the cluster within the Galactic
potential well, which has been modeled by using a three-component
axisymmetric analytic model. {It turned out that, at least during the last 3 Gyr, M71 has been in a disk-like orbit confined within the Galactic disk. It therefore seems reasonable to suppose that M71 suffered a number of dynamical processes (e.g., with the dense surrounding environment, the Milky Way spiral arms, various molecular clouds) that made it lose an amount of mass significantly larger than what expected for the majority of Galactic GCs, which are on halo-like orbits.}
We re-determined the gravitational center
and density profile of M71 by using resolved star counts from a
wide-field near-infrared catalog obtained with WIRCam at the
CFHT. This allowed us to minimize the impact of the large and
differential reddening affecting the system. With respect to the
values quoted in the literature (which have been determined from
optical data), we found the cluster centre to be located almost
$6\arcsec$ to the west, a $\sim 50\%$ larger core and half-mass radii.
Finally, we used a simplified analytical approach to take into account
mass-loss processes due to stellar and dynamical evolution, and thus
estimate the initial cluster mass, finding that the system likely was
one order of magnitude more massive than its current value.

As discussed in Sect. \ref{intro}, M71 is known to harbor a rich
population of X-ray sources \citep{elsner08}, in a number that exceeds
the predictions based on the values of its mass and its collision
parameter $\Gamma$ \citep{huang10}. Since this latter depends on the
cluster central luminosity density and core radius ($\Gamma
\propto\rho_0^{1.5} r_c^2$; \citealp{verbunt87, huang10}), we have
re-evaluated it by using the newly determined structural parameters.
By adopting the central surface brightness quoted in \citet{harris96}
and equation (4) in \citet{djorg93}, we found $\log\rho_0=2.60$ (in
units of $L_\odot$/pc$^{-3}$).  From this quantity and the value of
$r_c$ here determined, the resulting value of $\Gamma$ is about half
the one quoted in \citet{huang10}, and the discrepancy in terms of the
expected number of X-ray sources aggravates.  Instead, the much larger
initial mass here estimated for the system would be able to naturally
account for the currently observed X-ray population, thus reinforcing
the hypothesis that M71 lost a large fraction of stars during its
orbit.  An accurate investigation of the possible presence of tidal
tails around the cluster would be important to confirm such a
significant mass-loss from the system. However, this is currently
hampered by the large differential reddening affecting this region of
the sky, and a wide-field infrared observations are urged to shed
light on this issue.

\section{Acknowledgement}

We warmly thank the referee, whose useful comments improved the quality of the manuscripts.


\clearpage


\end{document}